\documentclass[sigconf]{acmart}
\usepackage{subfig}
\usepackage{natbib}
\usepackage{amsmath,amsfonts}
\usepackage{algorithmic}
\usepackage{graphicx}
\usepackage{textcomp}
\usepackage{footmisc}
\usepackage{xcolor}
\usepackage{multirow}
\usepackage{amsmath}
\usepackage{balance}       
\usepackage{graphics}      
\usepackage[T1]{fontenc}   
\usepackage{txfonts}
\usepackage{mathptmx}
\usepackage{booktabs}
\usepackage{textcomp}
\PassOptionsToPackage{hyphens}{url}
\usepackage{microtype}        
\usepackage{ccicons}          
\usepackage{graphicx}
\usepackage{caption}
\usepackage{float}
\usepackage{array}
\usepackage{makecell}
\usepackage{multirow}
\usepackage{wrapfig}
\usepackage{amsmath,amssymb,amsfonts}
\usepackage{algorithmic}
\usepackage{graphicx}
\usepackage{textcomp}
\usepackage{amsmath,amssymb,amsfonts}
\usepackage{algorithmic}
\usepackage{graphicx}
\usepackage{textcomp}
\usepackage[ampersand]{easylist}
\usepackage{amssymb}
\usepackage{multicol}
\usepackage{multirow}
\usepackage{tabularx}
\usepackage{rotating}
\usepackage[roman]{parnotes}
\usepackage[many]{tcolorbox}
\usepackage{soul}
\usepackage{comment}
\usepackage{booktabs}
\usepackage{colortbl}
\usepackage{verbatim}
\usepackage{array}
\usepackage[inline]{enumitem}
\usepackage[export]{adjustbox}
\usepackage{pifont}
\usepackage{mathtools} 
\usepackage{amsmath}
\usepackage{hyphenat}

\DeclareTextFontCommand{\mytexttt}{\ttfamily\hyphenchar\font=45\relax}
\AtBeginDocument{%
  \providecommand\BibTeX{{%
    \normalfont B\kern-0.5em{\scshape i\kern-0.25em b}\kern-0.8em\TeX}}}

\newif\ifdraft
\drafttrue 

\usepackage{xspace}

\usepackage{xcolor} 
\definecolor{darkgreen}{rgb}{0.05,0.5,0.05}

\newcommand{\gh}{GitHub\xspace}
\newcommand{\CE}{OE\xspace} 

\usepackage{fontawesome}

\definecolor{boxcolor}{RGB}{238, 223, 204} %
\DeclareRobustCommand{\mybox}[2][gray!10]{%
\begin{tcolorbox}[   
        breakable,
        left=0pt,
        right=0pt,
        top=0pt,
        bottom=0pt,
        colback=#1,
        colframe=black,
        width=\dimexpr\columnwidth\relax, 
        enlarge left by=0mm,
        boxsep=5pt,
        outer arc=4pt,
        boxrule=.5mm
        ]
        #2
\end{tcolorbox}
}

\copyrightyear{2022}
\acmYear{2022}
\setcopyright{acmcopyright}\acmConference[ICSE-SEIS'22]{Software Engineering in Society}{May 21--29, 2022}{Pittsburgh, PA, USA}
\acmBooktitle{Software Engineering in Society (ICSE-SEIS'22), May 21--29, 2022, Pittsburgh, PA, USA}
\acmPrice{15.00}
\acmDOI{10.1145/3510458.3513020}
\acmISBN{978-1-4503-9227-3/22/05}
\begin{document}

\title{Attracting and Retaining OSS Contributors with a Maintainer~Dashboard}


\author{Mariam Guizani}
\affiliation{%
  \institution{Oregon State University}
  \city{Corvallis}
  \state{Oregon}
  \country{USA}}
\email{guizanim@oregonstate.edu}

\author{Thomas Zimmermann}
\affiliation{%
  \institution{Microsoft Research}
  \city{Redmond}
  \state{Washington}
  \country{USA}}
\email{tzimmer@microsoft.com}

\author{Anita Sarma}
\affiliation{%
  \institution{Oregon State University}
  \city{Corvallis}
  \state{Oregon}
  \country{USA}}
\email{anita.sarma@oregonstate.edu}

\author{Denae Ford}
\affiliation{%
  \institution{Microsoft Research}
  \city{Redmond}
  \state{Washington}
  \country{USA}}
\email{denae@microsoft.com}

\renewcommand{\shortauthors}{Guizani et al.}

\begin{abstract}
Tools and artifacts produced by open source software (OSS) have been woven into the foundation of the technology industry.
To keep this foundation intact, the open source community needs to actively invest in sustainable approaches to bring in new contributors and nurture  existing ones.
We take a first step at this by collaboratively designing a maintainer dashboard that provides recommendations on how to \textit{attract} and \textit{retain} open source contributors. For example, by highlighting project goals (e.g., a social good cause) to attract diverse contributors and mechanisms to acknowledge (e.g., a ``rising contributor'' badge) existing contributors. Next, we conduct a project-specific evaluation with maintainers to better understand use cases in which this tool will be most helpful at supporting their plans for growth. 
From analyzing feedback, we find recommendations to be useful at signaling projects as welcoming and providing gentle nudges for maintainers to proactively recognize emerging contributors. However, there are complexities to consider when designing recommendations such as the project current development state (e.g., deadlines, milestones, refactoring) and governance model.
Finally, we distill our findings to share what the future of recommendations in open source looks like and how to make these recommendations most meaningful over time.
\end{abstract}

\maketitle




\section*{Lay Abstract}
Open Source Software (OSS) plays an important role in the development and maintenance of software products that are widely deployed in different domains from computer science to astrophysics and cutting edge medicines research. Chances are there is an open source project for anyone to contribute to. With the recent deployment of the popular Linux open source project  on Mars even the sky is no limit. However, OSS projects largely depend on volunteers and  attracting, retaining, and keeping contributors engaged is a severe challenge. In this paper, we present the design and evaluation of a dashboard to support community managers, such as maintainers, to track and acknowledge newcomers' contributions. With the support of tools such as ours, maintainers will be better prepared to attract and retain their emerging community.

\section{Introduction}
Despite the ubiquity of open source in our technology infrastructure and products, OSS projects struggle with a variety of challenges that can impact project health and sustainability. Research has found OSS to be a challenging ecosystem to navigate, both for  newcomers~\cite{steinmacher2015social, steinmacher2014attracting} and for experienced OSS contributors in large mature organizations~\cite{guizani2021long}. 

We scope our study to helping \textbf{attract and retain newcomers} as this group is essential in ensuring a constant flow of contributors and a sustainable community. Newcomers have been found to face particularly challenging barriers. They often struggle with even finding a task to work on~\cite{steinmacher2015social, steinmacher2014attracting}. Even after newcomers have contributed to the project, retaining them is nontrivial. 
In some projects, as high as 80\% of these contributors do not transition to long term contributors~\cite{steinmacher2013newcomers}. 

The challenges with attracting and retaining newcomers are not just unfortunate for newcomers. A lack of newcomers can also hurt the health and sustainability of OSS projects. It is thus important to implement proactive measures to help make OSS projects more welcoming for newcomers to join and stay. 

Previous research has focused on solutions to help newcomers, for example, through mentoring~\cite{fagerholm2014role, steinmacher2021being,fagerholm2014onboarding} or an information platform that overcomes the common newcomers' barriers~\cite{steinmacher2016overcoming}. In this paper, we focus on the other end of the spectrum by \textbf{empowering maintainers to improve their project and community health} for newcomers. OSS maintainers are at the forefront of leading the project and ensuring that the project's vision stays alive. We designed a first recommendation dashboard prototype to specifically help maintainers be intentional about attracting and retaining newcomers. 
The core \textbf{research question (RQ)} for this paper is as follows:

\mybox{\faQuestionCircle~\textbf{How useful are project and community growth recommendations for OSS projects?}}

To maximize impact we involved different stakeholders. (1) For the design, we collaborated with \emph{Open Source Experts} (\CE{}s), who regularly communicate with OSS participants such us maintainers and contributors, to ensure that we are designing a prototype that is in alignment with the information flow maintainers are familiar with. (2) We evaluated our dashboard with 8 \emph{OSS maintainers}.

This paper provides a first prototype of a maintainers' dashboard that focuses, in its first iteration, on attracting and retaining newcomers. We hope that our prototype and evaluation will help pave the way to incorporating recommendations into \gh and help maintainers and project leaders be proactive about their project and community growth.
\section{Background}
The health of an open source project depends on a healthy contributor base. Projects therefore need to attract, train, and retain contributors. However, attracting and retaining new contributors is a challenge. 
Newcomers to a project face a wide variety of challenges~\cite{steinmacher2015social,Steinmacher.Chaves.ea_2014}, that encompass the very first steps a contributor needs to take such as finding a starter task and a lack of response~\cite{steinmacher2015systematic, steinmacher2013newcomers, balali2020recommending}. Even when newcomers are able to make a contribution, transitioning them to become long term contributors is nontrivial~\cite{lee2017understanding, pinto2016more}. 
In fact, some projects end up losing 80\% of their newcomers~\cite{steinmacher2013newcomers} and almost half of contributors of a project contribute only once~\cite{pinto2016more}.
Even experienced contributors continue to face challenges in a mature OSS organization~\cite{guizani2021long}.

The challenges that contributors face include not finding a good task to start with~\cite{steinmacher2015social}, a lack of mentors ~\cite{balali2020recommending}, not being recognized for one's work ~\cite{guizani2021long}, a mismatch between contributors' motivation in joining a project and the project goals~\cite{gerosa2021shifting}, as well as a mismatch in career goals and expectations~\cite{trinkenreich2020hidden}. Scaffolding newcomer learning or encouraging contributors is not an easy task, as community leaders (e.g., maintainers) have limited time to attend to both the project and community needs \cite{pinto2018challenges}. 

Past research has identified strategies to overcome some of these challenges. Several works have researched mechanisms to help newcomers start to contribute. For example, newcomer-friendly issue labels that explicitly highlight starter tasks can help newcomers make their first contribution~\cite{steinmacher2015social, ChaossInclusiveIssueLabels, opensourceguide}. Santos et al.~\cite{Santos2021_MSR} provided an approach to automatically identify the skills needed for an OSS tasks, which could then be used to label issues. On the other hand, Huang et al.~\cite{huang2021leaving} have shown that certain project topics, especially those that relate to social good help in attracting a diverse set of contributors. Other works have shown the benefit of badges as an attractor for projects~\cite{trockman2018adding, qiu2019signals}. For example, code quality badges that signal project quality~\cite{spence2002signaling} can help create positive impression among contributors~\cite{qiu2019signals}. 

Here we draw on these research-evidenced strategies (e.g., highlighting starter tasks, explicitly stating project goals, and recognizing continued contributions) when generating our recommendations.

\section{Approach}
\begin{figure}[h]
  \includegraphics[width=\linewidth]{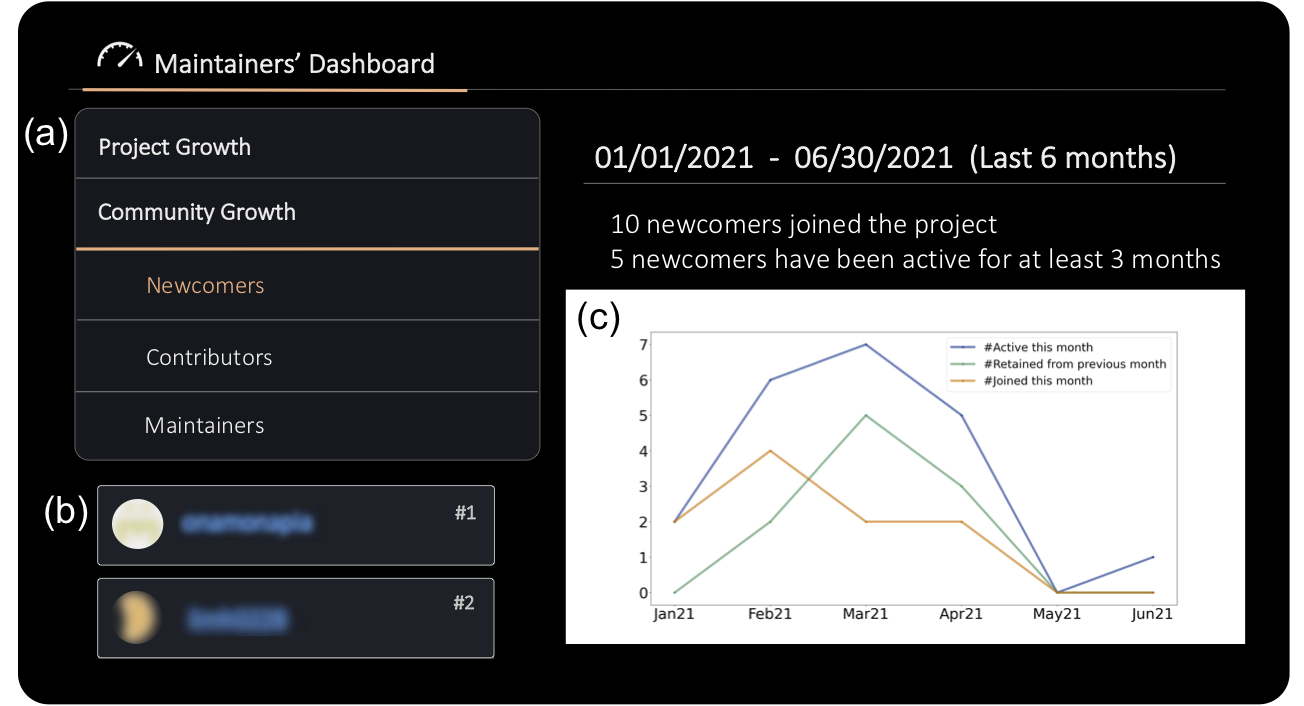}
  \caption{Community growth recommendation showing (a) the navigation menu, (b)  collapsed view of the recommended newcomers (see Figure~\ref{fig:risingContribExample}), and (c) graph showing the joining, activity, and retention trends of newcomers. See supplemental material for the complete UI as presented to maintainers~\cite{supplemental-materials}.}
  \label{fig:newcomergrowth}
\end{figure}
\subsection{Dashboard Design}
To support the two core challenges of attracting and retaining new contributors, we designed a dashboard to propose the different recommendations as we wanted to provide background information of why this recommendation was being proposed.
We designed the dashboard with the \gh interface so that it mirrored the interactions where maintainers manage other contributor activities. The dashboard was created in Figma~\cite{figma} and is reflected in Figure~\ref{fig:newcomergrowth}. Our supplemental materials includes an example of the full dashboard~\cite{supplemental-materials}.

 To ensure our dashboard was tailored to the most relevant use cases, we conducted multiple co-design sessions with our \CE{s} to prioritize the recommendations to be included in our dashboard. In these sessions, we used prior literature and \CE{s} insight from OSS research studies to determine project growth recommendations (i.e., those that impact project artifacts such as files, issues, PRs) and a community growth recommendation (i.e., those that affect contributor profiles such as individual badges). For the project growth recommendations (attract), we used issue labels of \textsc{good-first-issue} and \textsc{first-timers-only} as well as adding a topical tag that reflects the type of impact this project will have (see Figure~\ref{fig:ProjectGrowthRecomendation}(b)).
For the community growth recommendation (retain), the corresponding recommendation is to add the \textsc{rising contributor} badge.

\subsection{Project-Specific Dashboard Evaluation}
To understand the usefulness of these recommendations, we conducted project-specific interviews with maintainers from open source for social good (OSS4SG) projects. 
We used OSS4SG projects (and their maintainers) for our evaluation, as prior work found that OSS4G projects attract contributors who are interested in a broader impact~\cite{huang2021leaving}---which is well-aligned with our project growth recommendation. 
Additionally, as these projects also tend to attract contributors from minority groups, they provided a good opportunity for us to recruit participants from diverse backgrounds.

As interviews were customized for each maintainer, each interview took multiple steps to prepare.
First, once the maintainer interview was confirmed, we mined the project activity in the last 6 months (01/2021-06/2021), to generate newcomers activity graphs (see Figure~\ref{fig:newcomergrowth}(c)) and rising contributors to recommend (see Figure~\ref{fig:risingContribExample}). 
Second, we detected the presence (or lack of) newcomer-friendly issues by analyzing the issue tracker data to identify the state of newcomer-friendly issues in the project in order to recommend adding newcomer-friendly issue labels (see Figure~\ref{fig:ProjectGrowthRecomendation}(a)).
Third, we identified the OSS4SG goal(s) that were aligned with the project to recommend the accurate project growth badge(s) ~\cite{huang2021leaving}.

To recruit participants, we randomly selected 71 projects from the list of 434 OSS4SG projects~\cite{huang2021leaving} where the project still existed on \gh and where we could run the community growth recommendation. For instance, if a project did not have newcomers activity in the last six months it was not selected. We contacted our participants via email. We sent out recruitment emails to the maintainers of the 71 projects. The email included a brief description of the study, the compensation, and a recruitment intake form to complete. The recruitment survey provided the participants with a consent form with an opt-in/ opt-out option, a demographics survey and a link to schedule an interview time. 
We received eight responses resulting in a response rate of 11.3\%. Maintainers reported being from 5 different countries, identified as men and had an OSS experience ranging between 2 and 8 years.
Interviews ranged from 30 minutes to 45 minutes, after which, we thanked our participants and compensated them with a \$50 gift card. 

Each interview was audio and screen recorded for transcript and analysis purposes. Using grounded theory, we qualitatively analyzed the data of the eight interviews and summarized our findings on the usefulness of these recommendations, the lessons learned and ways to improve. We also performed a readout and shared our findings with the \CE{s}.

\section{Maintainers' Dashboard}
The OSS4SG Maintainers we interviewed found the dashboard prototype to be useful in managing the project and providing a reminder \textit{``that there's more to [helping the project] than [code]''(M8)}. The two following subsections summarize our findings on the usefulness of project growth recommendations (for attracting newcomers) and the community growth recommendation (for retaining newcomers).
\subsection{Project Growth: Attracting Newcomers}
For project growth, the dashboard uses \textit{two recommendations} to help attract newcomers: newcomer-friendly issue labels, and OSS4SG goal(s) project tags. 
\subsubsection{Newcomer-Friendly Issue Labels}

The dashboard analyzes the issue tracker data to identify: (1) the percent of issues that are labeled with a newcomer friendly label (Figure~\ref{fig:ProjectGrowthRecomendation}(a)) by string-matching against the set of newcomer-friendly labels in “MunGell / awesome-for-beginners” \gh repository~\cite{MunGell}. This  list  is  curated  from 187 repositories in 22 programming languages. Additionally, the dashboard detects if the issue labels the project currently uses include newcomer friendly labels used by other OSS projects~\cite{MunGell}. Based on the above data, the dashboard suggests adding newcomer friendly label(s) such us ``good first issue''(see Figure~\ref{fig:ProjectGrowthRecomendation}(a)).

The maintainers we interviewed had the following feedback regarding this dashboard recommendation.

\textbf{Signals a welcoming project...} Maintainers found that the ``good first issue label'' recommendation would increase the chances of newcomers in finding appropriate issues that \textit{``helps people getting into the project''}(M2), but also acts as a reminder for maintainers: \textit{``It is something I've thought about [but] haven't done, and probably nudge me to do it''}(M1). Maintainer M5 shared that these issue labels can  signal \textit{``a viewpoint that we want people to contribute, that obviously lifts the spirits of the project''}(M5) and shows that \textit{``there is help in the community''}(M5).

\textbf{...but depends on project characteristics.} Maintainers reported that the usefulness of the recommendation (issue label) depends on the development state of the project (e.g., milestones, refactoring). For instance maintainer M2 pointed out that \textit{``it's tricky...because we are having pretty strict timelines and milestones and prioritization''}. To help better accommodate their needs, M2 suggested a ``remind me later'' feature to reintroduce this recommendation at a later time. 

\subsubsection{OSS4SG Goals}
\begin{figure}%
    \centering
    \subfloat[\centering Example of newcomer-friendly issue label recommendation.]{{\includegraphics[width=3.8cm]{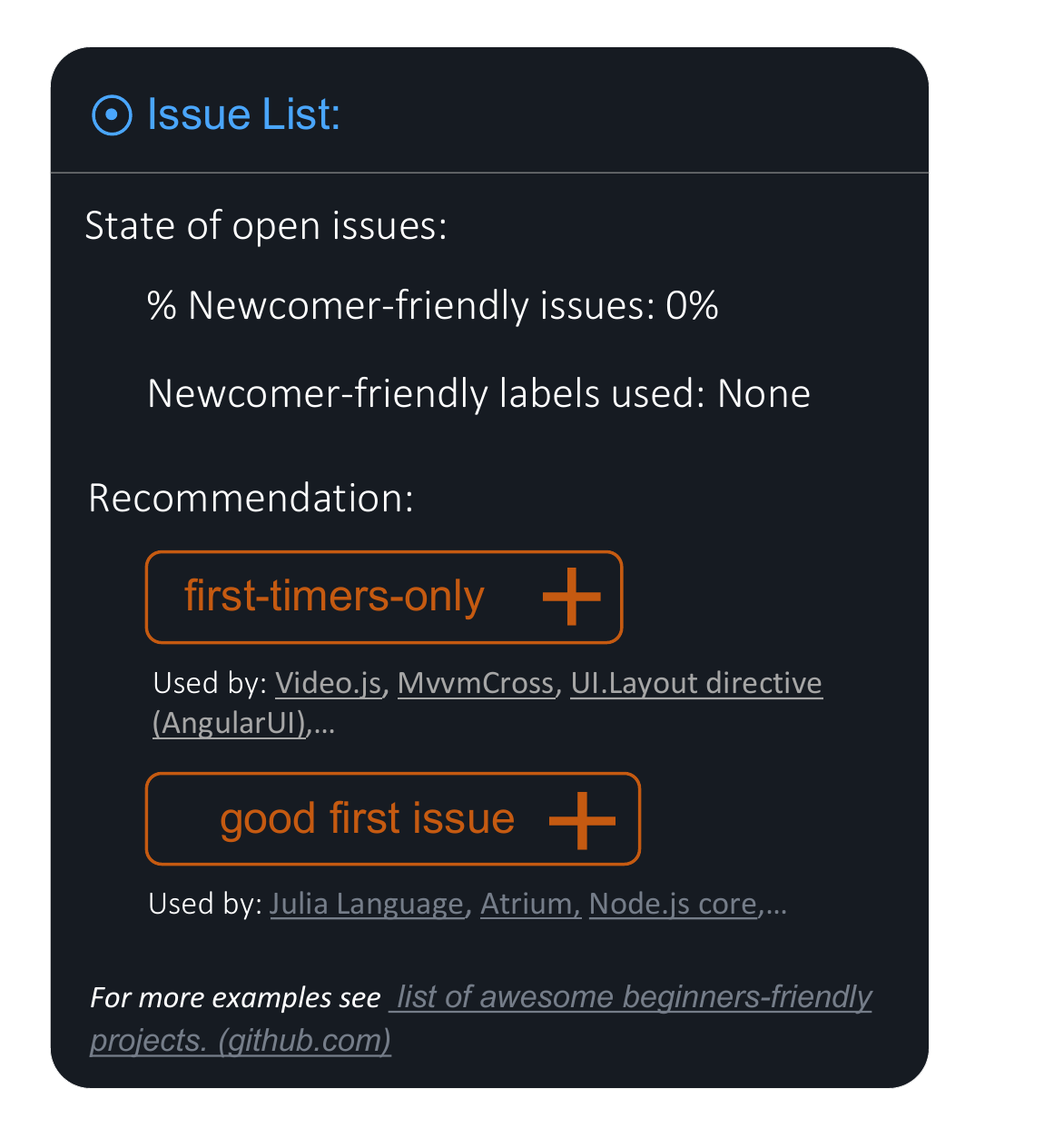} }}%
    \qquad
    \subfloat[\centering Example OSS4SG goal(s) project tags recommendation.]{{\includegraphics[width=3.8cm]{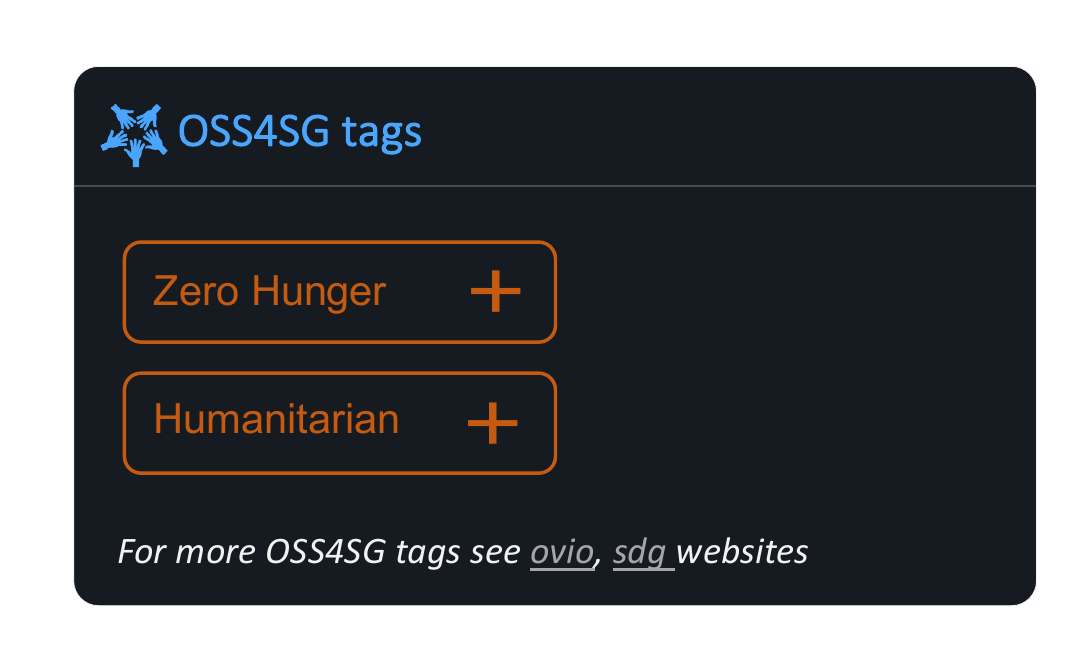} }}%
    \caption{Example of project growth recommendations.}%
    \vspace{-0.4cm}
    \label{fig:ProjectGrowthRecomendation}%
\end{figure}

We used the approach from ~\citet{huang2021leaving} to collect OSS4SG projects' goals. The dashboard recommended projects that are about ``social good'' to signal their goals through \textsc{readme.md} badges to help attract contributors (see Figure~\ref{fig:ProjectGrowthRecomendation}(b)). %
Maintainers found this recommendation can help their project in the following ways.

\textbf{Increases visibility...} Maintainers felt \textit{``those meta tags allow it [project] to be more contextually meaningful (M4)''} and \textit{``make it [project] more more visible''(M4)}, which was considered ``\textit{important, particularly, for young projects}'' (M6). %

\textbf{...and attracts diverse contributors and users.} Maintainers felt this recommendation would help \textit{``engage a broader spectrum''(M1)} of contributors, especially from \textit{``traditionally underrepresented people in computer science''(M1)}. M8 found that this could be useful to not only attract contributors but also users by indicating

\textit{``there's this project out there that can be interesting and useful to you. Never mind contributing to it, but use it''}. Maintainers also shared that this can emphasize that \textit{``working in the social good ones [projects] doesn't diminish the experience for the developers''(M1)}.
\vspace{2em}
\subsection{Community Growth: Retaining Newcomers}

\begin{wrapfigure}{l}{0.23\textwidth}
\begin{center}
    \centering
    \vspace{-0.5cm}%
    \includegraphics[width=0.25\textwidth]{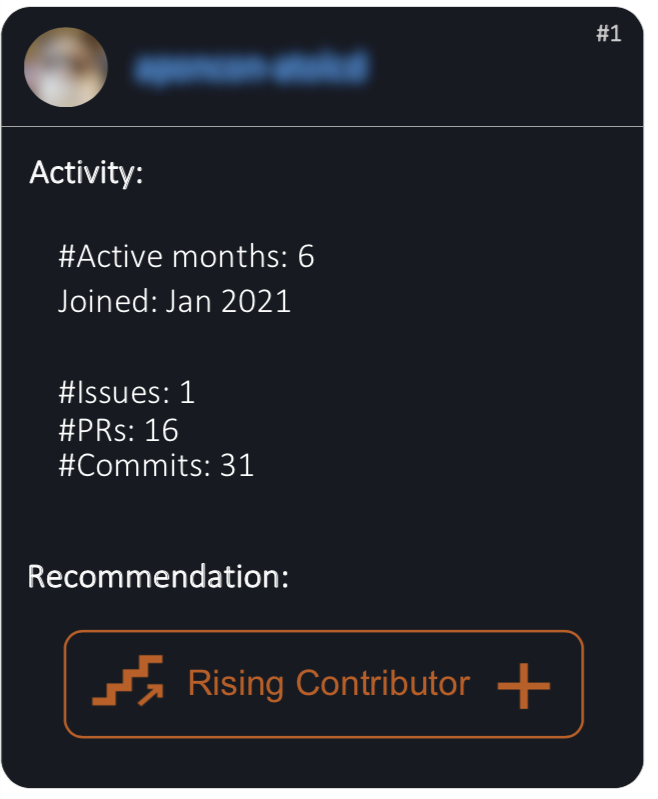}
    \caption{Example of ``rising contributor'' recommendation. }
    \vspace{-0.4cm}%
    \label{fig:risingContribExample}%
\end{center}
\end{wrapfigure}
For the community growth recommendation we focused on newcomers' retention. The example here focuses on recognizing active newcomers with a ``rising contributor'' badge. 

We identify ``rising contributors'' by analyzing the newcomers who joined the project in the last six months prior to the interviews and had not contributed to the project before. Using this data we generated newcomers' joining, activity and retention trends (see Figure~\ref{fig:newcomergrowth}(c)). We used consistent contribution as a proxy to gauge if a newcomer is a ``rising contributor''. If a newcomer contributed at least in the 3 out of the last 6 months they were identified as a ``rising contributor" (see Figure~\ref{fig:risingContribExample}). See supplemental for the detailed approach~\cite{supplemental-materials}.

We used the \gh API to mine OSS4SG software repositories and mined the commit, issue and pull request logs from 434 out of the 437 list of OSS4SG projects collected by \citet{huang2021leaving}. The three omitted projects no longer existed on \gh. 
Following are the four core feedback from maintainers about this badge (Figure~\ref{fig:risingContribExample}) and its source (Figure~\ref{fig:newcomergrowth}(c)).

\textbf{Recognizes contributors...} Maintainers shared that \textit{``it's nice to acknowledge people's work if we have a system that helps do that''(M8)} and that recognizing newcomers is\textit{ ``something that [they] wanted to know how to encourage better''(M6)}. 

More specifically they felt the badges benefit contributors as \textit{``a way to gain credibility''(M1)} and encourage them to continue contributing. M6 said: \textit{``...what we can do to recognize people's contributions better, I'm all for, and I'd be very happy for GitHub to implement''}.

\textbf{...but depends on the contribution model.} Maintainers pointed out that contributors' compensation attribute needs to be taken into consideration, especially as the open source model is now shifting to a hybrid model with paid and unpaid contributors coexisting. For example, M2 recognized the benefit of the badge, but \textit{``I know for a fact that they get paid to contribute to the project. So it won't really affect their motivation... I also think this would [be] pretty cool for us in specific circumstances like if we have someone who is voluntarily working on the project''}. To better accommodate a hybrid OSS model, maintainers suggested adding a filter by team membership to be able to recognize and acknowledge the work of unpaid contributors. Maintainer M6 whose core contributors are full-time employees thought of a different usage for the recommendation: as a hiring mechanism to detect potential new employees where \textit{``if someone had been contributing to our open-source project for six months to a year and consistently doing that, we probably would have hired them by then''}.

\textbf{Helps understand contributor activities...} 
Maintainers appreciated the activity information presented in Figure~\ref{fig:risingContribExample}. M4 \textit{``like(ed) that the activity is noted''} as a way \textit{``to see what kind of contributions the contributors are making''(M2)} in a more comprehensive way than just the commits as \textit{``a lot can happen particularly with new developers before commit'' (M1)}. M2 found the information particularly helpful \textit{``because it gives a fast overview about who's coding for the project [and] who's creating issues''}.

\textbf{... and overall activity.} The trend graph (Figure~\ref{fig:newcomergrowth}(c))  helped maintainers assess \textit{``that the community is growing'' (M2)}. In particular, understanding how certain events affect the rate of joining, activity and retention of contributors. M3 explained how \textit{``This would give an interesting set of data points...if you're having some hackathon event...or conference''}. M3 also shared the difficulty in retaining newcomers post-event and how it is very common to \textit{``have a spike and only a handful remain''}.

\section{Concluding Remarks}
In this work, we took a first step in providing a systematic mechanism through which OSS community leaders (e.g., maintainers) can be proactive about their community's health and sustainability. We designed and evaluated a dashboard that provides recommendations for maintainers to help them attract (e.g., advertise project goals) and retain (e.g., recognize contributors) newcomers to their project.

We worked closely with \CE{}s, who understand the project practices and communicate regularly with OSS participants. We then evaluated our dashboard with 8 maintainers who shared their feedback on the usefulness of the tool and ways to improve.

One key feedback when creating recommendations was the need to take into consideration the project's contribution model. 
With the open source ecosystem increasingly using hybrid models of contribution, different governance models exist.  While some projects are driven by a community of volunteers, others are company-driven with paid employees being the main contributors. M4 explained:\textit{ ``because this project in specific was company sponsored...Most of the things that work here [in the project], don't work elsewhere''.}

Other project characteristics such as the project's current development cycle (e.g., upcoming release, project restructuring) can also impact it's needs and it's ability to support newcomers.
As future work, we plan to continue to collaborate with the \CE{}s to expand our dashboard design to accommodate these different project characteristics. For example, maintainers requested recommendations for retaining core contributors and for detecting potential new maintainers.  

To summarize, open source software is a key digital infrastructure that drives many of our products and a venue for workforce development. However, current project maintainers are often resource-constrained to scaffold newcomers learning, which is not only detrimental to the project's health, but also to society. Therefore, as researchers we need to identify mechanisms to help maintainers attract and retain (new) contributors. As different OSS projects have different characteristics and needs, it is important to continue to ``listen'' to maintainers to better understand what tools and support they need. Maintainers' listening tours~\cite{GitHubAllinWebsite, AllInOSSGitHubProject} and industry conferences~\cite{OSSsummit, LinuxPlumbers} are great venues to foster industry-academia collaborations.

\begin{acks}
We thank all our survey respondents and our interview participants for their time and insight. We also thank our project stakeholders Bobby Dresser, Grace Vorreuter, and Evangeline Liu from the GitHub Communities Team,  Mala Kumar from the Social Impact Team and others for sharing their insight and expertise. Mariam Guizani performed this work during a summer internship  at  Microsoft  Research  in  the  Software  Analysis  and  Intelligence  Team  (\url{http://aka.ms/saintes}).  
This work is partially supported by the National Science Foundation (grants 1815486, 1901031).
\end{acks}




\bibliographystyle{ACM-Reference-Format}
\bibliography{biblio}


\end{document}